\documentclass[journal]{IEEEtran}

\usepackage{amsmath, bm, hyperref, authblk, nomencl, lipsum, chngcntr, blindtext}
\usepackage{tikz}
\usepackage[utf8]{inputenc}
\usepackage{pbox}
\usepackage{cite}
\usepackage[english]{babel}
\usepackage{array,colortbl,multirow,multicol,booktabs,ctable}

\usepackage{placeins}
\usepackage{forest}
\usetikzlibrary{arrows,decorations.pathmorphing,backgrounds,fit,positioning,shapes.symbols,chains, shapes,decorations}
\usepackage{csquotes}
\usepackage{comment}
\usepackage{color}
\usepackage{amsfonts}
\usepackage{array}
\usetikzlibrary{arrows.meta}
\usetikzlibrary{shapes,arrows}
\usetikzlibrary{automata, positioning}

\usepackage[makeroom]{cancel} 

\usetikzlibrary{arrows,decorations.pathmorphing,backgrounds,fit,positioning,shapes.symbols,chains, shapes,decorations}

\title{Stochastic and Distributionally Robust Load Ensemble Control}
\author{Ali Hassan, Robert Mieth, Deepjyoti Deka and Yury Dvorkin}	
\date{} %Leaves date blank 
\usepackage{amsthm}
\theoremstyle{plain}
\newtheorem{theorem}{Theorem}

\theoremstyle{definition}

\usepackage{placeins}
\newcommand{\subparagraph}{}
\usepackage{titlesec}
\usepackage{caption}
\theoremstyle{definition}

\theoremstyle{remark}
\newtheorem*{remark}{Remark}

\DeclareMathOperator{\KL}{KL}

\begin{document}			% Starts document/ends 
\clearpage
\thispagestyle{empty}
\maketitle

\begin{abstract}
Demand response (DR) programs aim to engage distributed demand-side resources in providing ancillary services for electric power systems. Previously, aggregated thermostatically controlled loads (TCLs) have been demonstrated as a technically viable and economically valuable provider of such services that can effectively compete with conventional generation resources in reducing load peaks and smoothing demand fluctuations. Yet, to provide these services at scale, a large number of TCLs must be accurately aggregated and operated in sync. This paper describes a Markov Decision Process (MDP) that aggregates and models an ensemble of TCLs. Using the MDP framework, we propose to internalize the exogenous uncertain dynamics of TCLs by means of stochastic and distributionally robust optimization. First, under mild assumptions on the underlying uncertainty, we derive analytical stochastic and distributionally robust control policies for dispatching a given TCL ensemble. Second, we further relax these mild assumptions to allow for a more delicate treatment of uncertainty, which leads to distributionally robust MDP formulations with moment- and Wasserstein-based ambiguity sets that can be efficiently solved numerically.  The case study compares the analytical and numerical control policies using a simulated ensemble of 1,000 air conditioners.
\end{abstract}

\begin{IEEEkeywords}
 Markov Decision Process (MDP), Linearly Solvable MDP, Distributionally Robust MDP, Thermostatically Controlled Loads, Uncertainty
\end{IEEEkeywords}

\section{Introduction} \label{Sec:introduction}
Thermal inertia of cooling and heating systems enables temporarily adjusting power consumption of thermostatically controlled loads (TCLs) without compromising their primary functions \cite{battle_group_2019, chertkov2017ensemble}. 
In the presence of constantly growing volatility and uncertainty of nodal power injections in electric power distribution systems caused by the integration of distributed energy resources (DERs), \textcolor{black}{thermal flexibility} of TCLs is a valuable control resource, \cite{battle_group_2019}. 
The ongoing expansion of grid-edge communication infrastructure also allows for designing demand response (DR) programs that enroll distributed small-scale flexible loads to provide various grid support services, both at the transmission and distribution levels. The Federal Energy Regulatory Commission (FERC) reports an increasing trend of DR program participation in the wholesale markets with a growth of 3\% from 2016 to 2017, to a total of 27,541 MW \cite{ferc_report}. To a large extent, this participation is enabled by aggregators that operate a large portfolio of similar devices \cite{lakshmanan2016impact} (called an ensemble) and act as mediators between grid operating entities, e.g. distribution system operators (DSOs), and individual flexible loads. The efficiency of these DR programs depends on the ability of aggregators to accurately model and control their ensembles.

TCLs, such as air conditioners, refrigerators or electric heaters, have a cycling pattern of energy consumption, i.e. they switch between on and off states given some user-defined thresholds (e.g. preferred temperature bands). This property allows to model TCL ensembles of an unlimited, or sufficiently large, size as a discrete-time, discrete-space Markov Process (MP) with relatively high accuracy. \textcolor{black}{Their power consumption can then be optimized using the Markov Decision Process (MDP) framework \cite{AC_TCL_Mathieu,Meyn_MDP,Chertkov_MDP,TCL_networks_Misha,MDP_Emiliano,Hassan_TCL,MDP_Turitsyn,MDP_Meyn,MDP_Meyn2}.}
The MP approach exploits the on/off switching behavior of TCLs and discretizes the ensemble dynamics into a finite number of states with each possible transition between these states characterized by a state-dependent probability. 
By capturing these transitions and their probabilities, the MP characterizes the interplay between the TCL temperature settings and electrical consumption based on external parameters (e.g. quality of refrigerator insulation, volume of air-conditioned space). 

In \cite{AC_TCL_Mathieu}, the authors show that the necessary parameters to construct such a MP representation can be obtained either from TCL electrical measurements or system temperature observations. The MP in \cite{AC_TCL_Mathieu} then employs a model predictive control strategy to achieve a desired consumption trajectory of the ensemble, thus allowing for dispatching TCLs like a virtual energy storage device. \textcolor{black}{Similarly, \cite{TCLs_Vrakopoulou,Callaway_TCL} developed methods to represent and dispatch TCL ensembles as virtual storage devices for providing regulation reserve.} In the context of DR aggregators, the desired load trajectory is the optimal trade-off between increasing the payoff of the aggregator and reducing comfort levels of TCL users, e.g. discomfort caused by deviations from their temperature settings. \textcolor{black}{By penalizing deviations from user-defined TCL settings, the MDP in \cite{Chertkov_MDP,Meyn_MDP} provides a tractable description of the TCL optimization by leveraging dynamic programming. MDP-based DR frameworks similar to \cite{AC_TCL_Mathieu,Chertkov_MDP,Meyn_MDP} can also accommodate network constraints to account for AC power flow and voltage limits in the distribution system \cite{TCL_networks_Misha,MDP_Emiliano}, as well as to mitigate the uncertainty of PV generation resources \cite{Hassan_TCL}. In addition, \cite{MDP_Turitsyn} consider the effect of fluctuating electricity prices on various types of controllable loads and derive a price-taking control strategy. Unlike \cite{TCL_networks_Misha,Meyn_MDP,MDP_Emiliano, MDP_Turitsyn,Hassan_TCL, Chertkov_MDP, AC_TCL_Mathieu, TCLs_Vrakopoulou,Callaway_TCL}, which assume that TCLs are operated in a centralize manner,  \cite{MDP_Meyn,MDP_Meyn2} develop a decentralized Markovian control strategy for an individual TCL resource to provide ancillary services to the power grid. %Notably, the authors in \cite{MDP_Turitsyn,MDP_Meyn} ignore network constraints in their formulations.
}

While \cite{AC_TCL_Mathieu,Meyn_MDP,Chertkov_MDP,TCL_networks_Misha,MDP_Emiliano,Hassan_TCL,MDP_Turitsyn,MDP_Meyn,MDP_Meyn2} demonstrate the usefulness of the MDP framework for dispatching TCL ensembles, they assume perfect knowledge of the ensemble transitions and their probabilities.
In practice, however, these model parameters are unknown and must be inferred from historical data. 
As available data on TCL ensembles is finite and potentially noisy, the true values of these model parameters remain unknown. This paper robustifies the MDP-based optimization of a risk-averse DR aggregator against uncertainty in the transition probabilities, thus 
generalizing the MDP models in \cite{Meyn_MDP,Chertkov_MDP,TCL_networks_Misha,Hassan_TCL}.
Leveraging methods of stochastic and distributionally robust optimization, we derive analytical and numerical methods to endogenously model uncertain transition probabilities and explore their potential effects on the optimal dispatch of TCL ensembles.

Parameter uncertainty arising from the inability to accurately estimate transition probabilities of the MP has been shown to significantly distort the outcomes of MDP solutions \cite{mannor2007bias}. The most common methods to overcome this caveat include percentile criteria \cite{delage2010percentile}, Kullback-Leibler divergence bounds \cite{nilim2004robustness}, nested uncertainty sets \cite{xu2010distributionally} or confidence regions using historical MDP performance metrics \cite{wiesemann2013robust}. This paper exploits an alternative approach and aims to internalize statistical information about the uncertainty on transition probabilities into the MDP optimization. Specifically, we explore how a mildly restrictive assumption enables a reformulation of the MDP optimization for TCLs as a linearly-solvable MDP (LS-MDP) \cite{LSMDP_todorov}. Using this LS-MDP framework and building on the previous work in \cite{Chertkov_MDP,TCL_networks_Misha,Hassan_TCL}, this paper  accounts for the transition probability uncertainty in the MDP optimization under different statistical assumptions summarized in Table~\ref{table_methodology_overview}. First, we use stochastic and distributionally robust optimization to derive analytical (closed-loop) control policies for the TCL ensembles under the assumption that the transition probability uncertainty is normally distributed, either with known or ambiguous distribution parameters. However, this assumption may still lead to unnecessarily erroneous TCL dispatch decisions. Second, we overcome the need for the normally distributed assumption, by introducing a moment-based ambiguity set into the MDP optimization that does not assume any distribution and only requires knowledge about first- and second-order moments. Although this approach does not result in a closed-form optimal control policy, we demonstrate that the MDP optimization under these assumptions can be solved efficiently with off-the-shelf solvers. To overcome the requirement on accurately computing the moments, we introduce a Wasserstein probability distance, \cite{wass_esfahani,Wasserstein_Gao}, in the distributionally robust MDP optimization and derive a computationally tractable reformulation. \textcolor{black}{Unlike the moment-based approach, the Wasserstein allows to capture all distributions within a pre-defined radius from a given nominal distribution, which can be drawn from empirical data, thus reducing data requirements needed to obtain a distributionally robust solution. Furthermore, the value of this radius can be used by decision-makers as a tuning parameter that allows for adjusting the solution conservatism.} To demonstrate and compare the performance of the presented analytical and numerical approaches, we conduct comprehensive numerical experiments on a TCL ensemble consisting of air conditioners.

\newcolumntype{M}[1]{>{\centering\arraybackslash}m{#1}}
\begin{table}[t]
  \centering 
  \caption{\textcolor{black}{Overview of the existing and proposed methods}}
  \label{table_methodology_overview}
  \begin{tabular}{M{2.6cm} c >{\raggedright}m{3cm} l}
    \toprule[.1em]
    \textbf{Method} & \textbf{Eq.} &\textbf{Uncertainty on transition probability} & \textbf{Solution} \\
    \midrule[.1em]
    Previous work, \cite{Chertkov_MDP,TCL_networks_Misha,Hassan_TCL} & \eqref{base_mdp} & None & Analytical \\
    \midrule
    Stochastic & \eqref{stochastic_final} & Normally distributed & \multirow{3}{*}{Analytical} \\
    \addlinespace
    Distributionally robust & \eqref{DRO_final} & Normally distributed with ambiguous parameters & \\
    \midrule
    Moment-based distributionally robust & \eqref{moment_final} & Any distribution with constraints on moments & \multirow{4}{*}{Numerical} \\
    \addlinespace
    Wasserstein-based distributionally robust & \eqref{wass_3} & Any distribution within a fixed distance of empirical distribution & \\
    \bottomrule
  \end{tabular}
\end{table}

\section{MDP for TCL Ensembles}

Building on our prior work in \cite{Chertkov_MDP,TCL_networks_Misha,Hassan_TCL}, we represent a homogeneous ensemble of sufficiently many TCLs as a discrete-time, discrete-space MDP. From the perspective of the DR aggregator, the optimization problem for operating the TCL ensemble is:
\begin{subequations}
\begin{align}
&\underset{\substack{\rho,\mathcal{P}_t}}{\text{min}} \ \mathbb{E}_{\rho}
\sum_{t \in \mathcal{T}} \! \sum_{\alpha \in \mathcal{A}} \big(-U_{t+1}^{\alpha} + \sum_{\beta \in \mathcal{A}} \gamma \log\! \frac{\mathcal{P}_{t}^{\alpha\beta}}{\overline{\mathcal{P}}_{}^{\alpha\beta}}\big) \label{MDP:obj} \hspace{-5cm} && \\
  \text{s.t.} \quad &\rho_{t+1}^{\alpha} = \sum_{\beta \in \mathcal{A}} \mathcal{P}_{t}^{\alpha\beta} \rho_{t}^{\beta}, && \forall \alpha \in \mathcal{A}, t \in \mathcal{T} %\backslash |N_\mathcal{T}|
  \label{MDP_evol} \\
  % & p_{t,b}= \sum_{\alpha \in \mathcal{A}} p_b^{\alpha} \rho_{t,b}^{\alpha}, && \forall t \in \mathcal{T}, b \in \mathcal{N} \label{mdp_injP1} \\
  &\sum_{\alpha \in \mathcal{A}} \mathcal{P}_{t}^{\alpha \beta} = 1,  && \forall \beta \in \mathcal{A}, t \in \mathcal{T} \label{mdp_integrality} 
  %&\sum_{\alpha} \mathcal{P}_{i,t}^{\alpha\beta} = 1
\end{align}%
\label{base_mdp}%
\end{subequations}%
\noindent \textcolor{black}{where $\rho \in \mathbb{R}^{n}$, $n=|\mathcal{A}|$, is a vector with entries $\rho_{t+1}^{\alpha} \geq 0$ and $\rho_{t}^{\beta} \geq 0$ representing the probabilities that the TCL ensemble is in states $\alpha,\beta\in\mathcal{A}$ at times $t+1$ and $t$, respectively, $\mathcal{A}$ is the set of all possible states, and operator $\mathbb{E}_{\rho}$ denotes the expectation over $\rho$.}
Set $\mathcal{A}$ is obtained by discretizing the range of aggregated power consumption of the ensemble given the operating range of each TCL \cite{Chertkov_MDP}. Probabilities $\rho_{t+1}^{\alpha}$ and $\rho_{t}^{\beta}$ are related via the transition probability matrix $\mathcal{P}_{t} \in \mathbb{R}^{n\text{x}n}$, with $n=|\mathcal{A}|$, and where entry $\mathcal{P}_{t}^{\alpha\beta}$ of matrix $\mathcal{P}_{t}$ characterizes the probability of the transition of the TCL ensemble from state $\beta$ at time $t$ to state $\alpha$ at time $t+1$. 
Note that the TCL ensemble can also remain in the same state such that $\alpha = \beta$. 
On the other hand,  matrix $\mathcal{\overline{P}} \in \mathbb{R}^{n\text{x}n}$ with entries $\mathcal{\overline{P}}^{\alpha\beta}$ represents the default transition probability, i.e. the steady state behavior of the ensemble without any control actions of the aggregator. 
\textcolor{black}{Additionally, internal control actions such as user-defined settings and their on-demand adjustments can still be applied to the individual TCLs in the ensemble, which will modify and will be reflected in default transitions and the probability matrix. (The inability to perfectly forecast these internal control actions introduce the uncertainty that we deal with in Sections~\ref{Analytical_formulation}-\ref{Numerical Control}.)}
\textcolor{black}{
In the following, we treat the vector $\rho$ and matrix $\mathcal{P}_{t}$ as decision variables, which can be achieved by suitable TCL control actions \cite{AC_TCL_Mathieu}.} In contrast, entries $\mathcal{\overline{P}}^{\alpha\beta}$ of matrix $\mathcal{\overline{P}}$ are treated as parameters of the MDP optimization in~\eqref{base_mdp}. Although matrix $\mathcal{\overline{P}}$ is modeled as time-independent, unlike $\mathcal{P}_t$, this modeling choice can be revisited, if sufficient historical data about the TCL ensemble is available. \textcolor{black}{As more empirical data on the TCL dispatch is collected over time, the more temporal fidelity can be achieved in representing default transitions. All methods to account for the uncertainty presented below will hold if $\mathcal{\overline{P}}$ is modeled as time-dependent.}

Eq.~\eqref{MDP:obj} is the objective function of the aggregator that operates the TCL ensemble and tries to maximize its expected utility $U^{\alpha}_{t+1}$ at future state $\alpha$ at time $t+1$ and to minimize the discomfort cost of the TCL ensemble, which is modeled as the logarithmic difference between the uncontrolled transitions of the TCL ensemble ($\mathcal{\overline{P}}^{\alpha \beta}$) and the resulting transition probabilities due to the control decisions of the aggregator ($\mathcal{P}_{t}^{\alpha\beta}$). \footnote{\textcolor{black}{The discomfort cost of TCLs can be interpreted as a change in their temperature settings from user-defined comfort/convenience levels, e.g. for freezers, air-conditioners, hot-water tanks, heat pumps, and swimming pool pumps.}} This discomfort cost in the second term of \eqref{MDP:obj} can be interpreted as the Kullback-Leibler (KL) divergence weighed by cost penalty $\gamma$, \cite{KL_book}. 
The KL divergence is widely used for modeling discrepancies in discrete- and continuous-time series, \cite{KL_time_series}, and makes it possible to derive closed-form optimal control policies. 
\textcolor{black}{Parameter $\gamma$ can influence the KL divergence and thus encourage or discourage deviations from the default behavior of the TCL ensemble.}
\textcolor{black}{Furthermore, if $\mathcal{\overline{P}}^{\alpha \beta}=0$, i.e. a transition from state $\beta$ to $\alpha$ has not been observed in the past, the model in \eqref{base_mdp} restricts $\mathcal{P}_{t}^{\alpha\beta}=0$ and excludes such transitions when optimizing it for the rest of the values.}
Eq.~\eqref{MDP_evol} describes the temporal evolution of the TCL ensemble from time $t$ to $t+1$ over time horizon $\mathcal{T}$. 
Eq.~\eqref{mdp_integrality} imposes the integrality constraint on the transition decisions optimized by the aggregator such that their total probability is equal to one. \textcolor{black}{After solving \eqref{base_mdp}, the active power ($p_{t}$) consumed by the TCL ensemble can be computed using decisions $\rho^{\beta}_{t}$ and rated active power $p^{\beta,rated}$ at each state as $p_{t}= \sum_{\beta \in \mathcal{A}} p^{\beta} \rho_{t}^{\beta,rated},\forall t \in \mathcal{T}$.} \textcolor{black}{Since \eqref{base_mdp} is formulated for a discrete-time MP, the resulting dispatch does not capture power fluctuations between discrete time instances. However, since the TCL ensemble is assumed to be sufficiently large, random fluctuations of TCL loads neutralize one another at the ensemble level, \cite{PhysRevE.101.022115}. Furthermore, the residual effects of such fluctuations between discrete time instances can be mitigated if one uses a more fine-grained temporal resolution. However, the latter may increase computing times. }

Our prior work in \cite{TCL_networks_Misha,Hassan_TCL} shows that the optimization in \eqref{base_mdp} is a LS-MDP as introduced by \cite{LSMDP_todorov}. The LS-MDP has no explicit actions, is controlled by modifying a predefined (uncontrolled) probability distribution over subsequent states as modeled by decisions \textcolor{black}{$\mathcal{P}_{t}^{\alpha\beta}$}. 
The optimal policy obtained from \eqref{base_mdp} is a next-state distribution, which minimizes the accumulated state costs of the agent traversing state space $\mathcal A$, while minimizing the divergence cost between the controlled ($\mathcal{P}_{t}^{\alpha\beta}$) and uncontrolled ($\overline{\mathcal{P}}^{\alpha\beta}$) probability distributions. This optimal policy can be computed as:

\begin{theorem} \label{theorem0} \normalfont
Let \eqref{base_mdp} model a TCL ensemble as a LS-MDP. Then the optimal control policy is:
\begin{align}
&\mathcal{P}_{t}^{\alpha \beta} = \frac{\overline{\mathcal{P}}^{\alpha \beta}z_{t+1}^{\alpha}}{\sum_{\alpha \in \mathcal{A}}\overline{\mathcal{P}}^{\alpha \beta}z_{t+1}^{\alpha}},  
\end{align}
where $z_{t+1}^{\alpha}=\text{exp}(-\varphi_{t+1}^{\alpha}/\gamma)$ and value function $\varphi_{t+1}^{\alpha}$ is defined as $\varphi_{t+1}^{\alpha}$=$-U_{t+1}^{\alpha}-\gamma\text{log}\big( \sum_{\textcolor{black}{\upsilon}\in\mathcal{A}}\text{exp}\big( \frac{-\varphi_{t+2}^{\textcolor{black}{\upsilon}}}{\gamma}\big)\overline{\mathcal{P}}^{\textcolor{black}{\upsilon} \alpha} \big),$ \textcolor{black}{where $\textcolor{black}{\upsilon} \in \mathcal{A}$ is a state at time $t+2$.}
\end{theorem}
\begin{proof} 
See proof in Appendix \ref{sec:exp_zero_mean}.
\end{proof}

Theorem \ref{theorem0} implies that computing the optimal control policy depends on the uncontrolled transition probability ($\mathcal{\overline{P}}^{\alpha \beta}$) and the value function of the next state ($\varphi_{t+1}^{\alpha}$). 
However, this requires the default transition probabilities to be perfectly known, which does not hold in real-world applications, where the TCL ensemble is subject to unknown external influences and uncertain human behavior. We model this parameter uncertainty by representing default transition probabilities $\overline{\mathcal{P}}^{\alpha\beta}$ as random variables $\overline{\boldsymbol{\mathcal{P}}}^{\alpha\beta}$, indicated by the \textbf{bold} font. As summarized in Table~\ref{table_methodology_overview}, we derive and study methods to internalize $\overline{\boldsymbol{\mathcal{P}}}^{\alpha\beta}$ in the optimal MDP control policy using different assumptions and statistical information on $\overline{\boldsymbol{\mathcal{P}}}^{\alpha\beta}$.

\begin{remark}
\textcolor{black}{Although the MDP in \eqref{base_mdp} is developed for a homogenous TCL ensemble, it can be extended to modeling heterogenous TCL ensembles. For instance, one can classify TCL loads in a given heterogenous ensemble and represent it as a set of homogeneous subensembles. Then, each subensemble can be operated separately using the proposed MDP framework. Similarly, the models proposed in Sections~\ref{Analytical_formulation} and \ref{Numerical Control} can be extended to operating heterogenous TCL ensembles.}
\end{remark} 
\section{Analytical Control Policies} \label{Analytical_formulation}

The standard MDP formulation in \eqref{base_mdp} allows the derivation of a closed-form optimal control policy as shown by Theorem~\ref{theorem0}. The goal of this section is to show that this useful property can be maintained if $\overline{\boldsymbol{\mathcal{P}}}^{\alpha\beta}$ is normally distributed.

\subsection{Stochastic Formulation}
Assume that $\overline{\boldsymbol{\mathcal{P}}}^{\alpha\beta}$ follows a normal distribution with mean $\overline{\mathcal{P}}^{\alpha\beta}$ and variance $\sigma^{2}$, i.e. $\overline{\boldsymbol{\mathcal{P}}}^{\alpha\beta}\sim N(\overline{\mathcal{P}}^{\alpha\beta},\sigma^{2})$. 
\textcolor{black}{The mean and variance can be calculated from a set of $N$ historical observations of $\overline{\boldsymbol{\mathcal{P}}}^{\alpha\beta}$ that can be retrieved by the aggregator from operating data of a given TCL ensemble\footnote{\textcolor{black}{We ensure $\sum_{\alpha\in\mathcal{A}}\overline{\boldsymbol{\mathcal{P}}}^{\alpha\beta}=1$. In other words, the probability of moving from present state $\beta$ to all possible next states $\alpha$ is equal to one}.}. We denote this set of observations as \textcolor{black}{$\{\overline{\mathcal{P}}^{\alpha\beta}_{j,obs}\}_{j\in N}$} and use it to infer distribution parameters such as empirical mean ($\overline{\mathcal{P}}^{\alpha\beta}$) and variance ($\sigma^{2}$) as follows:}
\begin{align}
&\hspace{-1mm} \overline{\mathcal{P}}^{\alpha\beta} =\! \frac{1}{N} \sum_{j \in N}\textcolor{black}{\overline{\mathcal{P}}^{\alpha\beta}_{j,obs}},\ \sigma^{2} = \frac{1}{N-1}\! \sum_{j \in N} (\textcolor{black}{\overline{\mathcal{P}}^{\alpha\beta}_{j,obs}}-{\overline{\mathcal{P}}^{\alpha\beta}})^{2}
\end{align}
Then, we reformulate \eqref{base_mdp} as:
\begin{subequations}
\begin{align}
& \hspace{-0.8cm} \underset{\substack{\rho,\mathcal{P}}}{\text{min}}\ O^{E}:= \mathbb{E}_{{\overline{\boldsymbol{\mathcal{P}}}^{\alpha\beta}}} \mathbb{E}_{\rho}
\!\!\sum_{t \in \mathcal{T}}\! \! \sum_{\alpha \in \mathcal{A}}\!\! \bigg(\!\!-U_{t+1}^{\alpha}\! +\!\! \sum_{\beta \in \mathcal{A}}\! \gamma \log\! \frac{\mathcal{P}_{t}^{\alpha\beta}}{\overline{\boldsymbol{\mathcal{P}}}^{\alpha\beta}}\bigg) \label{uncertain_1} \\
\text{s.t.} \quad &\text{Eq. } \eqref{MDP_evol}-\eqref{mdp_integrality},
\end{align}%
\label{uncertain_init}%
\end{subequations}%
\noindent where $\mathbb{E}_{{\overline{\boldsymbol{\mathcal{P}}}^{\alpha\beta}}}$ denotes the expectation over ${\overline{\boldsymbol{\mathcal{P}}}^{\alpha\beta}}$ and $\mathbb{E}_{\rho}$ is identical to~\eqref{MDP:obj}. Eq.~\eqref{uncertain_1} can further be simplified as:
\begin{align}
&\!\!\!O^{E}\!\!=\!\mathbb{E}_{\rho}\!\! \sum_{t \in \mathcal{T}} \! \! \sum_{\alpha \in \mathcal{A}}\!\!\! \Big\{\!\!-\!U_{t+1}^{\alpha}\! +\!\gamma\!\! \sum_{\beta \in \mathcal{A}}\!\! \big( \log\! \mathcal{P}_{t}^{\alpha\beta}\!\!-\!\!\mathbb{E}_{{\overline{\boldsymbol{\mathcal{P}}}^{\alpha\beta}}}\! [\log\!{\overline{\boldsymbol{\mathcal{P}}}^{\alpha\beta}} ] \big)\!\! \Big\} \label{uncertain_3}
\end{align}
\noindent where the last term can be approximated by the second-order Taylor expansion as, \cite[Eq. (17)]{expect_log}:
\begin{align}
\mathbb{E}_{{\overline{\boldsymbol{\mathcal{P}}}^{\alpha\beta}}} [\log ({\overline{\boldsymbol{\mathcal{P}}}^{\alpha\beta}})] &\approx \log\mathbb{E}_{{\overline{\boldsymbol{\mathcal{P}}}^{\alpha\beta}}}[\overline{\boldsymbol{\mathcal{P}}}^{\alpha\beta}] - \frac{\text{Var}({\overline{\boldsymbol{\mathcal{P}}}^{\alpha\beta}})}{2(\mathbb{E}_{{\overline{\boldsymbol{\mathcal{P}}}^{\alpha\beta}}}[{\overline{\boldsymbol{\mathcal{P}}}^{\alpha\beta}}])^2} \nonumber \\
 &= \log({\overline{\mathcal{P}}^{\alpha\beta}}) - \frac{\sigma^2}{2({\overline{\mathcal{P}}^{\alpha\beta}})^2}. \label{expec_3}
\end{align}
%Eq.~\eqref{expec_3} follows from the second-order Taylor expansion as done in \cite{expect_log}.
Given~\eqref{expec_3}, the optimization in~\eqref{uncertain_init} is rewritten as:
\begin{subequations}
\begin{align}
&\underset{\substack{\rho,\mathcal{P}}}{\text{min}}\ O^{E}:= \mathbb{E}_{\rho} \sum_{t \in \mathcal{T}} \sum_{\alpha \in \mathcal{A}}\! \Big\{\! -U_{t+1}^{\alpha} + \gamma \!\!\sum_{\beta \in \mathcal{A}}\! \Big(\log \frac{\mathcal{P}_{t}^{\alpha\beta}} {{\overline{\mathcal{P}}^{\alpha\beta}}} \nonumber \\
& \qquad \qquad \qquad + \frac{\sigma^2}{2({\overline{\mathcal{P}}^{\alpha\beta}})^2} \Big) \Big\} \label{uncer_obj_final}\\
\text{s.t.} \quad &\text{Eq. } \eqref{MDP_evol}-\eqref{mdp_integrality} \label{mdp_cons}
\end{align}%
\label{stochastic_final}%
\end{subequations}%
Given the stochastic formulation in~\eqref{stochastic_final}, we prove:

\begin{theorem} \label{theorem1} \normalfont
Let \eqref{stochastic_final} model a TCL ensemble as a LS-MDP with uncertain transition probabilities defined as $\overline{\boldsymbol{\mathcal{P}}}^{\alpha\beta}\sim N(\overline{\mathcal{P}}^{\alpha\beta},\sigma^{2})$. Then the optimal control policy is:
\begin{align}
&\mathcal{P}_{t}^{E} := \mathcal{P}_{t}^{\alpha \beta} = \frac{\overline{\mathcal{P}}^{\alpha \beta}z_{t+1}^{\alpha}\text{exp}\big(\frac{-\sigma^2}{2({\overline{\mathcal{P}}^{\alpha\beta}})^2}\big)}{\sum_{\alpha}\overline{\mathcal{P}}^{\alpha \beta}z_{t+1}^{\alpha}\text{exp}\big(\frac{-\sigma^2}{2({\overline{\mathcal{P}}^{\alpha\beta}})^2}\big)}, \label{P_e} 
\end{align}
where $z_{t+1}^{\alpha}\!=\!\text{exp}(-\varphi_{t+1}^{\alpha}/\gamma)$ and value function $\varphi_{t+1}^{\alpha}$ is defined as $\varphi_{t+1}^{\alpha}$=$-U_{t+1}^{\alpha}-\gamma\text{log}\big(\! \sum_{\textcolor{black}{\upsilon}\in\mathcal{A}}\text{exp}\big( \frac{-\varphi_{\textcolor{black}{t+2}}^{\textcolor{black}{\upsilon}}}{\gamma}\big)\overline{\mathcal{P}}^{\textcolor{black}{\upsilon} \alpha}$ $\text{exp}\big( \frac{-\sigma^2}{2({\overline{\mathcal{P}}^{\textcolor{black}{\upsilon}\alpha}})^2}  \big)  \big),$ \textcolor{black}{where $\textcolor{black}{\upsilon} \in \mathcal{A}$ is a state at time $t+2$.}
\end{theorem}
\begin{proof} 
See proof in Appendix \ref{sec:exp_zero_mean}.
\end{proof}

Similarly to Theorem~\ref{theorem0}, the optimal control policy obtained from Theorem \ref{theorem1} depends on the mean values of uncontrolled transition probabilities ($\overline{\mathcal{P}}^{\alpha\beta}$), the next-state value function ($\varphi_{t+1}^a$) and variance ($\sigma^2$). However, term $\text{exp}\Big(\frac{-\sigma^2}{2({\overline{\mathcal{P}}^{\alpha\beta}})^2}\Big)$ distinguishes the control policy in Theorem~\ref{theorem1} from Theorem~\ref{theorem0} and internalizes the  uncertainty on uncontrolled transition probabilities into the optimal control policy. Hence, the stochastic solution in Theorem \ref{theorem1} is anticipated to improve the optimal control policy formulated in Theorem \ref{theorem0} for an average performance of the TCL ensemble. 
However, Theorem~\ref{theorem1} still exploits the assumption that parameters of the uncertainty distribution, i.e. $\overline{\mathcal{P}}^{\alpha\beta}$ and $\sigma^{2}$, are perfectly known.

\subsection{Distributionally Robust Formulation} \label{sec:dro_simple}
To internalize potential parameter misestimation due to the finite number of available observations, we leverage distributionally robust optimization that allows for modeling the inferred distribution parameters via an ambiguity set. In this setting, the objective of the DR aggregator is to maximize their expected performance under the worst-case distribution of $\overline{\boldsymbol{\mathcal{P}}}^{\alpha\beta}$ drawn from a given ambiguity set denoted as $\mathbb{D}$:
\begin{subequations} \label{eq:DRO}
\begin{align} 
\begin{split}
&\underset{\substack{\rho,\mathcal{P}}}{\text{min}}\ O^{WC}:= \underset{\overline{\boldsymbol{\mathcal{P}}}\in\mathbb{D}}{\text{sup}} \mathbb{E}_{\rho}
\sum_{t \in \mathcal{T}} \sum_{\alpha \in \mathcal{A}} \Big\{-U_{t+1}^{\alpha} \\& \qquad \qquad \quad +\gamma \sum_{\beta \in \mathcal{A}} \Big( \log \frac{\mathcal{P}_{t}^{\alpha\beta}} {{\overline{\mathcal{P}}^{\alpha\beta}}} + \frac{\sigma^{2}}{2({\overline{\mathcal{P}}^{\alpha\beta}})^2} \Big) \Big\} \label{DRO_1}
\end{split}\\
\text{s.t.} \quad &\text{Eq. } \eqref{MDP_evol}-\eqref{mdp_integrality} \label{DRO_2},
\end{align}
%\label{DRO_obj}
\end{subequations}
The ambiguity set in~\eqref{eq:DRO} is defined as $\mathbb{D}= [ \underline{\Gamma}\! \leq \overline{\mathcal{P}}^{\alpha\beta}\!\!\! \leq \overline{\Gamma} ,\underline{\hat{\zeta}}\! \leq {\sigma}^2\! \leq \overline{\hat{\zeta}}]$, where $\underline{\Gamma}$, $\overline{\Gamma}$, $\underline{\hat{\zeta}}$ and $\overline{\hat{\zeta}}$ are confidence bounds on the empirical mean and variance. Since $\overline{\mathcal{P}}^{\alpha\beta}$ and ${\sigma}^2$ can be respectively modeled by $t$- and Chi-Square ($\mathcal{X}^2$) distributions \cite{chi_sq_dist}, we compute these bounds as:
\begin{align}
& \!\!\!\underline{\Gamma} = \overline{\mathcal{P}}^{\alpha\beta}\!\! - t_{(1-\varsigma/2)} \frac{\sigma}{\sqrt{N}} \text{ and } \ \overline{\Gamma} = \overline{\mathcal{P}}^{\alpha\beta}\!\! + t_{(1-\varsigma/2)} \frac{\sigma}{\sqrt{N}}, \label{eq:ambiguity_set_mean}\\
& \underline{\hat{\zeta}} = \frac{(N-1)\sigma^{2}}{\mathcal{X}^2_{(1-\xi)/2}} \text{ and } \ \overline{\hat{\zeta}} = \frac{(N-1)\sigma^{2}}{\mathcal{X}^2_{\xi/2}}, \label{eq:ambiguity_set_var}
\end{align}
where we denote $t_{(1-\varsigma/2)}$ in \eqref{eq:ambiguity_set_mean} as the $(1-\varsigma/2)$-quantile of the $t$-distribution and $\mathcal{X}^2_{\xi}$ in \eqref{eq:ambiguity_set_var} as the $\xi$-quantile of the Chi-Square distribution.
Given $\mathbb{D}$, the objective function in~\eqref{DRO_1} can be reformulated as:
\begin{align}
\begin{split}
& \underset{\overline{\boldsymbol{\mathcal{P}}}\in\mathbb{D}}{\text{sup}} \sum_{t \in \mathcal{T}} \!\!\sum_{\alpha \in \mathcal{A}}\! \Big\{\!\!-U_{t+1}^{\alpha}\! +\!\gamma\! \sum_{\beta \in \mathcal{A}}\!\! \Big(\!\! \log \frac{\mathcal{P}_{t}^{\alpha\beta}} {{\overline{\mathcal{P}}^{\alpha\beta}}} + \frac{{\sigma}^2}{2({\overline{\mathcal{P}}^{\alpha\beta}})^2} \Big)\! \Big\} \\&\! = \sum_{t \in \mathcal{T}}\! \sum_{\alpha \in \mathcal{A}}\!\! \Big\{\!\!-U_{t+1}^{\alpha} + \!\gamma\!\sum_{\beta \in \mathcal{A}}\!\! \Big( \log \frac{\mathcal{P}_{t}^{\alpha\beta}} {\underline{\Gamma}} + \frac{\overline{\hat{\zeta}}}{2(\underline{\Gamma})^2} \Big) \Big\}, \label{DR0_3}
\end{split}
\end{align}
leading to the following optimization problem:
\begin{subequations}
\begin{align}
\begin{split}
&\!\!\!\!\!\!\!\!\!\underset{\substack{\rho,\mathcal{P}}}{\text{min}}\ O^{WC}:= \mathbb{E}_{\rho}
\!\!\sum_{t \in \mathcal{T}}\!\! \sum_{\alpha \in \mathcal{A}}\!\! \Big\{\!-U_{t+1}^{\alpha} +\!\gamma\! \sum_{\beta \in \mathcal{A}}\!\! \Big( \log \frac{\mathcal{P}_{t}^{\alpha\beta}}{\underline{\Gamma}}\\
  & \qquad \qquad \qquad \qquad \qquad+ \frac{\overline{\hat{\zeta}}}{2(\underline{\Gamma})^2} \Big) \Big\} \label{chi_obj}
\end{split}\\
\text{s.t.} \quad & \text{Eq. } \eqref{MDP_evol}-\eqref{mdp_integrality}. \label{chi_const}
\end{align}%
\label{DRO_final}%
\end{subequations}%
Given the reformulation of \eqref{eq:DRO} presented in \eqref{DRO_final}, we prove:
\begin{theorem} \label{theorem2} \normalfont
Let \eqref{DRO_final} model a TCL ensemble as a LS-MDP with $\overline{\boldsymbol{\mathcal{P}}}^{\alpha\beta}\sim N(\overline{\mathcal{P}}^{\alpha\beta},\sigma^{2})$ and $\overline{\mathcal{P}}^{\alpha\beta}, \sigma^2 \in \mathbb{D}$, where $ \mathbb{D} = [ \underline{\Gamma} \leq \overline{\mathcal{P}}^{\alpha\beta} \leq \overline{\Gamma} ,\underline{\hat{\zeta}} \leq {\sigma}^2 \leq \overline{\hat{\zeta}}]$. Then the optimal control policy is:
\begin{align}
&\mathcal{P}_{t}^{WC} := \mathcal{P}_{t}^{\alpha \beta} = \frac{\underline{\Gamma}z_{t+1}^{\alpha}\text{exp}\big(\frac{-\overline{\hat{\zeta}}}{2(\underline{\Gamma})^2}\big)}{\sum_{\alpha}\underline{\Gamma}z_{t+1}^{\alpha}\text{exp}\big(\frac{-\overline{\hat{\zeta}}}{2(\underline{\Gamma})^2}\big)} \label{P_wc}
\end{align}%
where $z_{t+1}^{\alpha}=\text{exp}(-\varphi_{t+1}^{\alpha}/\gamma)$ and value function $\varphi_{t+1}^{\alpha}$ is defined as $\varphi_{t+1}^{\alpha}$=$-U_{t+1}^{\alpha}-\gamma\text{log}\big(\! \sum_{\textcolor{black}{\upsilon}\in\mathcal{A}}\text{exp}\big( \frac{-\varphi_{t+2}^{\textcolor{black}{\upsilon}}}{\gamma}\big)\underline{\Gamma}\text{exp}\big( \frac{-\hat{\zeta}}{2(\underline{\Gamma})^2}\!\big)\!  \big),$ \textcolor{black}{where $\textcolor{black}{\upsilon} \in \mathcal{A}$ is a state at time $t+2$.}
\end{theorem}

\begin{proof} 
See proof in Appendix \ref{sec:exp_zero_mean}.
\end{proof}

Similarly to Theorems~\ref{theorem0} and \ref{theorem1}, Theorem \ref{theorem2} computes the optimal control policy using the mean values of the default transition probabilities ($\overline{\mathcal P}^{\alpha\beta}$) and the next-state value function ($\varphi_{t+1}^\alpha$).
However, it additionally internalizes the information about set $\mathbb{D}$ and immunize the optimal control policy for the worst-case realization of distribution parameters drawn from this set. This overcomes the need to perfectly know distribution parameters as in Theorem~\ref{theorem1}, thus improving the goodness of fit between the LS-MDP model and empirical data.

\subsection{Hybrid Model}
Relative to the stochastic formulation in \eqref{stochastic_final}, the distributional robustness of \eqref{DRO_final} imposes additional conservatism on the optimal control policy, which may lead to a greater solution cost. To trade-off the robustness and cost performance of the optimal policy, we seek the hybrid formulation that can weigh the stochastic and distributionally robust formulations via parameter $\eta$:
\begin{subequations}
\begin{align}
\begin{split}
&\underset{\substack{\rho,\mathcal{P}}}{\text{min}}\ (1-\eta) O^{WC} + \eta O^{E} \label{hybrid_obj}
\end{split}\\
\text{s.t.} \quad & \text{Eq. } \eqref{MDP_evol}-\eqref{mdp_integrality} \label{const_hybrid}
\end{align}%
\label{hybrid_final}%
\end{subequations}%
where $0\leq \eta \leq 1$.

\begin{theorem} \label{theorem3} \normalfont
Let \eqref{hybrid_final} model a TCL ensemble as a LS-MDP with uncertainty defined as $\overline{\boldsymbol{\mathcal{P}}}^{\alpha\beta}\sim N(\overline{\mathcal{P}}^{\alpha\beta},\sigma^{2})$ and $\overline{\mathcal{P}}^{\alpha\beta}$,$\sigma^2 \in \mathbb{D}$, where $ \mathbb{D} = [ \underline{\Gamma} \leq \overline{\mathcal{P}}^{\alpha\beta} \leq \overline{\Gamma} ,\underline{\hat{\zeta}} \leq {\sigma}^2 \leq \overline{\hat{\zeta}}]$. Then the optimal control policy is given as:
\begin{align}
\begin{split}
&\mathcal{P}_{t}^{\alpha \beta} = (1-\eta) \mathcal{P}_{t}^{WC} + \eta \mathcal{P}_{t}^{E}
\end{split}
\end{align}
where $0\leq \eta \leq 1$ is a parameter characterizing risk tolerance of the aggregator and $\mathcal{P}_{t}^{E}$ and $\mathcal{P}_{t}^{WC}$ are given by \eqref{P_e} and~\eqref{P_wc}.
\end{theorem}

\begin{proof} 
See proof in Appendix \ref{sec:exp_zero_mean}.
\end{proof}
% \vspace{-1cm}

Theorem \ref{theorem3} yields the optimal control policy that balances the stochastic and distributionally robust models weighted by parameter $\eta$, which can be set by the DR aggregator based on its risk tolerance.

\section{Numerical Control Policies} \label{Numerical Control}
The analytical control policies derived in the previous Section~\ref{Analytical_formulation} assume that $\overline{\boldsymbol{\mathcal{P}}}^{\alpha\beta}$ is normally distributed, even if distribution parameters are not precisely known and drawn from the ambiguity set. However, these assumptions may still limit the performance and applicability of the analytical policies. This caveat motivates a further investigation of methods that allow for more generic control policies. 

\subsection{Moment-based Ambiguity Set} \label{section_moment}
Instead of assuming a specific (e.g. normal) uncertainty distribution, we define $\overline{\boldsymbol{\mathcal{P}}}^{\alpha\beta}$ solely in terms of its statistical moments (e.g. mean and variance).
In other words, this approach achieves distributional robustness by defining an ambiguity set that captures all distributions with statistical moments satisfying given confidence parameters. Hence, we redefine uncertainty set $\mathbb{D}$:
\begin{equation} 
\begin{split}
& \mathbb{D} := \{ \mathbb{P} \in \mathcal{M}(\mathbb{R}) | \mathbb{P} (W) = 1 :(\nu) , \\
& \qquad \qquad  -b \leq \mathbb{E}_{\overline{\boldsymbol{\mathcal{P}}}^{\alpha\beta} \sim \mu}[\overline{\boldsymbol{\mathcal{P}}}^{\alpha\beta}] - m \leq b :(\underline{\lambda},\overline{\lambda}), \\
& \qquad \qquad \qquad [\mathbb{E}_{\overline{\boldsymbol{\mathcal{P}}}^{\alpha\beta} \sim \mu}(\overline{\boldsymbol{\mathcal{P}}}^{\alpha\beta} - m)^{2}] \leq c \sigma^{2} :(\Lambda) \},
\label{moment_constraints}
\end{split}
\end{equation}
\textcolor{black}{where $\mathbb{E}_{\overline{\boldsymbol{\mathcal{P}}}^{\alpha\beta} \sim \mathbb{P}}$ is the expectation over empirical probability distribution $\mathbb{P}$ supported by samples $\{\overline{\mathcal{P}}^{\alpha\beta}_y\}_{y\in N}$, $\mathcal{M}$ is the set of all distributions, $W$ is the support set, and $m$ and $\sigma^2$ are the nominal mean and variance with confidence parameters $b$ and $c$.} Given the nominal values and confidence parameters, the uncertainty set in \eqref{moment_constraints} allows for the worst-case mean and variance be drawn from a range of values.
%\textcolor{orange}{[I think we should specify the difference to the previous amb. set here again for clarity, because, technically, before we also defined a range of means and variances to pick from.]}
%Considering a range of possible values for the worst-case mean and variance makes the set in \eqref{moment_constraints} more flexible relative to the ambiguity set used in Section~III-\ref{sec:dro_simple}, which treats the mean and variance as given parameter.
Note that in \eqref{moment_constraints} we introduce dual variables $\nu, \underline{\lambda},\overline{\lambda},$ and $\Lambda$ for each constraint, which are given after a colon. Given the ambiguity set in \eqref{moment_constraints}, we define the following optimization problem:

\begin{subequations}
\begin{align}
&\hspace{-1mm} \underset{\substack{\rho,\mathcal{P}}}{\text{min}} \ \underset{\mathbb{P}\in\mathbb{D}}{\text{sup}} \ \mathbb{E}_{\overline{\boldsymbol{\mathcal{P}}}^{\alpha\beta} \sim \mathbb{P}}\mathbb{E}_{\rho}
\!\sum_{t \in \mathcal{T}} \! \sum_{\alpha \in \mathcal{A}}\!\! \bigg(\!\!\!-U_{t+1}^{\alpha} + \!\gamma\!\!\sum_{\beta \in \mathcal{A}}\! \log\! \frac{\mathcal{P}_{t}^{\alpha\beta}}{\overline{\boldsymbol{\mathcal{P}}}^{\alpha\beta}}\bigg) \label{moment_obj} \\
&\text{s.t.} \quad \text{Eq. } \eqref{MDP_evol}-\eqref{mdp_integrality} \label{moment_cons}.
\end{align}
\label{moment_opt}
\end{subequations}
\allowdisplaybreaks
Solving \eqref{moment_opt} is challenging because the optimization is performed over infinite dimensional set $\mathbb{D}$. To the best of our knowledge, such problems cannot be solved analytically and there are also no efficient computational tools \cite{infinite_optimization}. However, one way to tackle such problems is to leverage convex duality theory that transforms the original problem over an infinite dimensional set into a dual problem over finite dimensional Lagrange multipliers with the same value as the original problem \cite{Rockafellar_duality,Convex_Optimization_in_Infinite,Shapiro_duality}. The duality approach in an infinite dimensional setting is developed by Rockafellar in \cite{Rockafellar_duality} and is based on pairing locally convex topological vector spaces. The requirement of the existence of a feasible interior point (Karush-Kuhn-Tucker point) for the implicit constraint set is relaxed to require only continuity of the optimal value function. After transforming the original problem to its dual form, we can use finite optimization computational tools to obtain a solution. Therefore, we take the dual of the inner maximization problem and reformulate the objective function \eqref{moment_obj} as follows:
\allowdisplaybreaks
\begin{subequations}
\begin{align} 
\begin{split}
&\underset{\substack{\underline{\lambda},\overline{\lambda},\Lambda,\nu}}{\text{min}} \mathbb{E}_{\rho} \sum_{t \in \mathcal{T}} \! \sum_{\alpha \in \mathcal{A}} \bigg\{-U_{t+1}^{\alpha} + \gamma\sum_{\beta \in \mathcal{A}} \Big[\log \mathcal{P}_{t}^{\alpha\beta}\\& + (b - m)\underline{\lambda}+ (b + m)\overline{\lambda} + c \sigma^{2}\Lambda + \nu \Big] \bigg\} 
\end{split} \\
\text{s.t.}& \nonumber \\
& \!\!\!\!\!\!\!\!\!\!\!(\overline{\lambda}\! -\! \underline{\lambda})\overline{\boldsymbol{\mathcal{P}}}^{\alpha\beta}\!\!\! +\! \Lambda (\overline{\boldsymbol{\mathcal{P}}}^{\alpha\beta}\!\!\!\!-\! m)^{2}\!\! + \! \nu\! \geq \!- \!\log\! \overline{\boldsymbol{\mathcal{P}}}^{\alpha\beta},\ \forall \overline{\mathcal{P}}^{\alpha\beta}\!\!\! \in\! W, \label{dual_cons_moment}
\end{align}%
\label{eq:dual_inner}%
\end{subequations}%
where $\{\underline{\lambda},\overline{\lambda},\Lambda \geq 0;\ \nu\!:\!\!\text{free}\}$ are dual variables defined for the constraints in ambiguity set $\mathbb{D}$ given by \eqref{moment_constraints}. Eq.~\eqref{eq:dual_inner} represents an upper bound of the inner maximization in \eqref{moment_opt} because \eqref{moment_obj} essentially maximizes over a convex function ($\sup-\log\overline{\boldsymbol{\mathcal{P}}}^{\alpha\beta}$). By substituting \eqref{eq:dual_inner} in \eqref{moment_opt}, we obtain the following single-level optimization problem:
\allowdisplaybreaks
\begin{subequations}
\begin{align}
\begin{split}
& \hspace{-0.5cm} \underset{\substack{\rho,\mathcal{P},\underline{\lambda},\overline{\lambda},\Lambda,\nu}}{\text{min}} \mathbb{E}_{\rho}
\sum_{t \in \mathcal{T}} \! \sum_{\alpha \in \mathcal{A}} \!\!\bigg\{-U_{t+1}^{\alpha} + \gamma\sum_{\beta \in \mathcal{A}} \Big[\log \mathcal{P}_{t}^{\alpha\beta}\\& + (b - m)\underline{\lambda}+ (b + m)\overline{\lambda} + c \sigma^{2}\Lambda + \nu \Big] \bigg\}\label{moment_f_obj}
\end{split}\\
\text{s.t.} \quad & \text{Eq. } \eqref{MDP_evol}-\eqref{mdp_integrality}\label{moment_f_c1}\\
& \text{Eq. } \eqref{dual_cons_moment}\label{moment_f_c2}.
\end{align}
\label{moment_final}
\end{subequations}
%\allowdisplaybreaks[0]

The optimization problem in \eqref{moment_final} can be solved numerically with off-the-shelf solvers by discretizing $W$ in \eqref{dual_cons_moment}. Note that relative to the analytical control policies developed in Section~\ref{Analytical_formulation}, \eqref{moment_final} yields a numerical solution, yet with optimality guarantees. Although this numerical solution is less generalizable than the analytical solutions, it is obtained under less restrictive assumptions on the underlying uncertainty, which is more suitable for practical needs and allows one to avoid unnecessary conservatism of the optimal solution.

\subsection{Wasserstein-based Ambiguity Set}
Although the moment-based ambiguity set in \eqref{moment_constraints} avoids assuming a particular distribution, it still restricts the first- and second-order moments within given ranges determined by the confidence parameters, which is shown to produce overly conservative solutions for certain problems \cite{Wasserstein_Gao}. Hence, to alleviate the need to invoke these restrictions, we define an ambiguity set using the Wasserstein metric, which makes it possible to immunize the optimal solution against any distribution that lies within fixed radius $\psi > 0$ around a given nominal distribution. Accordingly, we formulate this ambiguity set as:
\begin{align}
& \mathcal{C}_\tau := \{\mathbb{P} \in \mathcal{M}: W_{p}(\mathbb{P},\hat{\mathbb{P}}) \leq \psi \}, \label{eq:wass_ball} 
\end{align}
\textcolor{black}{where $W_{p}$ is the Wasserstein metric of order $p$ evaluating the distance between distribution $\mathbb{P}$ and nominal distribution $\hat{\mathbb{P}}$.} Given empirical distribution $\overline{\boldsymbol{\mathcal{P}}}_{t}^{\alpha\beta}$ based on observations \textcolor{black}{$\{\overline{\mathcal{P}}_{y}^{\alpha\beta}\}_{y\in N}$}, the nominal distribution in \eqref{eq:wass_ball} can be defined as $\textcolor{black}{\hat{\mathbb{P}}} = \frac{1}{N}\sum_{y \in N}\delta_{\overline{\mathcal{P}}^{\alpha\beta}_y}$, where $\delta_{\overline{\mathcal{P}}^{\alpha\beta}_y}$ is a Dirac distribution for $\overline{\boldsymbol{\mathcal{P}}}^{\alpha\beta}_y$.
Hence, the Wasserstein distance between \textcolor{black}{distributions $\mathbb{P}$ and $\hat{\mathbb{P}}$ defines the minimum cost of redistributing mass from $\mathbb{P}$ to $\hat{\mathbb{P}}$}. Hence, using \eqref{eq:wass_ball}, we can reformulate the distributionally robust objective function as follows:
\begin{align}
\begin{split}
\hspace{-2mm} & \underset{\substack{\rho,\mathcal{P}}}{\text{min}} \underset{\substack{\mathbb{P} \in \mathcal{C}_\tau}}{\text{sup}} \mathbb{E}_{\overline{\boldsymbol{\mathcal{P}}}^{\alpha\beta}} \mathbb{E}_{\rho}\!
\sum_{t \in \mathcal{T}}\!\! \bigg(\!\sum_{\alpha \in \mathcal{A}}\!\!\!-U_{t+1}^{\alpha} + \! \gamma\!\!\sum_{\alpha \in \mathcal{A}}\!\sum_{\beta \in \mathcal{A}}\! \log\! \frac{\mathcal{P}_{t}^{\alpha\beta}}{\overline{\boldsymbol{\mathcal{P}}}^{\alpha\beta}}\bigg). \end{split}\label{wass_obj}
\end{align}
Using Definition 3.1 and reformulation steps in Section 4.1 from \cite{wass_esfahani}, \eqref{wass_obj} can be reformulated as:
\begin{subequations}
\begin{align}
\begin{split}
&\underset{\substack{\rho,\mathcal{P}},\lambda,s}{\text{min}}\ \mathbb{E}_{\rho}
\sum_{t \in \mathcal{T}} \! \bigg\{\!\sum_{\alpha \in \mathcal{A}}\!-U_{t+1}^{\alpha} + \!\gamma\!\sum_{\beta \in \mathcal{A}}\bigg(\sum_{\alpha \in\mathcal{A}}\log \mathcal{P}_{t}^{\alpha\beta} \\& \qquad \qquad \qquad + \lambda \psi + \frac{1}{N} \sum_{y \in N} s_{y} \bigg)\bigg\} 
\end{split}\\
\text{s.t.}& \nonumber \\
\begin{split}
&\hspace{-5mm} \!\underset{\substack{\overline{\mathcal{P}}^{\alpha\beta,\text{min}}\! \leq \overline{\boldsymbol{\mathcal{P}}}^{\alpha\beta} \!\leq \overline{\mathcal{P}}^{\alpha\beta,\text{max}}\\\sum_{\alpha\in\mathcal{A}}\overline{\boldsymbol{\mathcal{P}}}^{\alpha\beta}=1}}{\text{sup}}\!\sum_{\alpha \in \mathcal{A}}\!\! \big\{\!\! -\!\log\! \overline{\boldsymbol{\mathcal{P}}}^{\alpha\beta}\!\!\! - \lambda |\overline{\boldsymbol{\mathcal{P}}}^{\alpha\beta}\!\!\! -\!\overline{\mathcal{P}}_{y}^{\alpha\beta}|\! \big\} \\[-6mm]&\qquad\qquad\qquad\qquad\qquad \ \leq\! s_{y}, \forall \beta \in \mathcal{A}, y \in N
\end{split}\label{wass_sup_cosnt} \\
&\text{Eq. } \eqref{MDP_evol}-\eqref{mdp_integrality},
\end{align}
\label{wass_3}
\end{subequations}
where $s_y$ is an auxiliary variable and range $[\overline{\mathcal{P}}^{\alpha\beta,\text{min}},\overline{\mathcal{P}}^{\alpha\beta,\text{max}}]$ defines the support for $\overline{\boldsymbol{\mathcal{P}}}^{\alpha\beta}$, where parameters $\overline{\mathcal{P}}^{\alpha\beta,\text{min}}$ and $\overline{\mathcal{P}}^{\alpha\beta,\text{max}}$ are drawn from observations \textcolor{black}{$\{\overline{\mathcal{P}}_{y}^{\alpha\beta}\}_{y\in N}$}. Similarly to the relationship between \eqref{moment_opt} and \eqref{eq:dual_inner}, \eqref{wass_3} represents an upper bound of \eqref{wass_obj} because it also maximizes over a convex function ($\sup-\log\overline{\boldsymbol{\mathcal{P}}}^{\alpha\beta}$). Since \eqref{wass_sup_cosnt} is convex, the supremum of \eqref{wass_sup_cosnt} can be obtained by an exhaustive search over extreme points. The extreme points are generated by the intersection of hyper-boxes, representing the range $[\overline{\mathcal{P}}^{\alpha\beta,\text{min}},\overline{\mathcal{P}}^{\alpha\beta,\text{max}}]$, and a hyper-plane, ensuring that the probability of moving from present state $\beta$ to all possible next states $\alpha$ is equal to one ($\sum_{\alpha\in\mathcal{A}}\overline{\boldsymbol{\mathcal{P}}}^{\alpha\beta}=1$). This allows us to solve \eqref{wass_3} using off-the-shelf solvers.

\section{Case Study} \label{Case Study}

\begin{figure}[!t]
\centering
\includegraphics[width=0.9\columnwidth]{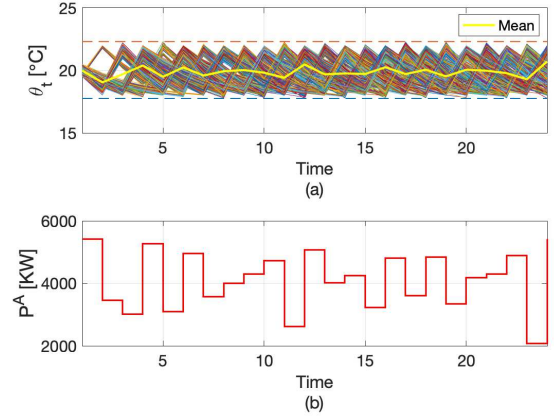}
\caption{(a) Temperature evolution of the ensemble with 1000 TCLs and (b) their aggregated power consumption.}
\label{fig:tcl}
\end{figure}
\begin{figure}[!t]
\centering
\includegraphics[height=4cm,width=.5\columnwidth]{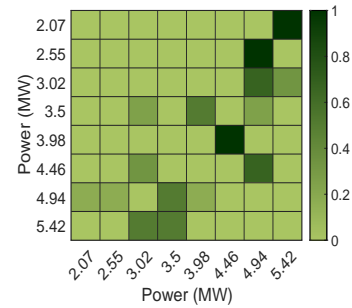}
\caption{Default transition probability matrix ($\overline{\mathcal{P}}^{\alpha\beta}$) with 8 states constructed from the power profile in Fig. 1(b), where the color density indicates the probability value in the sidebar.}
\label{fig:tran}
\end{figure}

The case study is carried out for a TCL ensemble with 1,000 residential air conditioner units. The discrete-time model for an individual residential air conditioner is based on \cite{AC_TCL_Mortensen,AC_TCL_Mathieu,AC_TCL_Samrat} and given as: 
\begin{align}
&\theta_{t+1} = \varrho \theta_{t} + (1-\varrho)(\theta^{a} - \aleph R P u_{t}) + \kappa_{t},  
\end{align}
where $\varrho = \text{exp}(-h/RC)$, $\theta_{t}$ represents the indoor temperature of the room, $\theta^{a}$ is the ambient temperature, $R$ is the thermal resistance, $C$ is the thermal capacitance, $P$ is the electrical power consumption, $u_{t}\in\{0,1\}$ determines whether the device is on or off, and $\aleph$ is the thermal efficiency. Parameter $\kappa_t$ represents noise, which is ignored in the construction of the MP, and instead is accounted for by randomizing the default transition probabilities and solve it using different methods as given in Table \ref{table_methodology_overview}. Fig.~\ref{fig:tcl} displays simulated temperature trajectories and the resulting aggregated power consumption. 
The aggregated power consumption is discretized into 8 states with uniform power intervals and the associated probability transitions are shown in Fig. \ref{fig:tran}. These transitions are defined as the default transition probabilities ($\overline{\mathcal{P}}$) in our models. Next, we generate 1,000 random samples representing the set of observations by varying default transition probabilities within 15\% of their nominal values in Fig. \ref{fig:tran}, while ensuring that the sum of probabilities remains equal to one. Then this set is used to estimate the empirical mean ($\overline{\mathcal{P}}^{\alpha \beta}$) and variance ($\sigma^2$) values. All simulations are performed using the Julia JuMP \cite{jump} package on an Intel Core i5 2.3 GHz processor with 8~GB of RAM and the Ipopt solver.

\subsection{Analytical Control Policies} \label{Case Study:Analytical}

\begin{table}[!t]
\centering
\scriptsize
\caption{Cost Performance of Analytical Control Policies.}
\vspace{3pt}
\begin{tabular}{ c|c|c|c|c|c }
\hline
\multicolumn{6}{c}{Solution Cost (Objective function), \$} \\
\hline
{$\gamma (\$)$}& {$\eta$=0.00} & {$\eta$=0.25} & {$\eta$=0.50}& {$\eta$=0.75} & {$\eta$=1.00}\\
\hline
0.05 & 2787.04 & 2786.63 & 2786.22 & 2785.81 & 2785.40 \\
0.10 & 2805.10 & 2804.43 & 2803.76 & 2803.09 & 2802.42 \\
1.00 & 2884.72 & 2879.16 & 2873.59 & 2867.99 & 2862.38 \\
\hline 
\end{tabular}

\label{table_hybrid}
\end{table}

\begin{figure}[!t]
\centering
\includegraphics[height=9cm,width=\columnwidth]{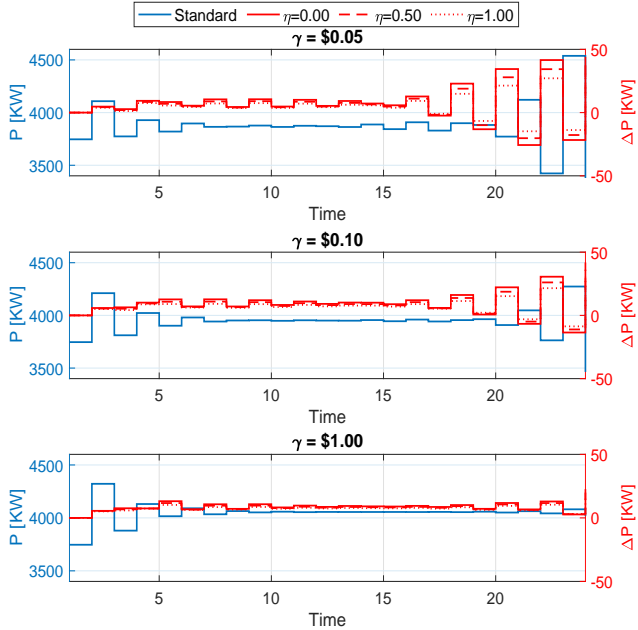}

\caption{Optimal power dispatch under the standard MDP in~\eqref{base_mdp} (blue) and the difference (denoted as $\Delta P$) in the power consumption under analytical stochastic and distributionally robust control policies for different values of cost penalty $\gamma$. The stochastic and distributionally robust policies are computed using the hybrid model in~\eqref{hybrid_final} with $\eta=1$ and $\eta=0$, respectively.}
\label{fig:hybrid}
\end{figure}

\begin{figure}[!t]
\centering
\includegraphics[height=9cm,width=\columnwidth]{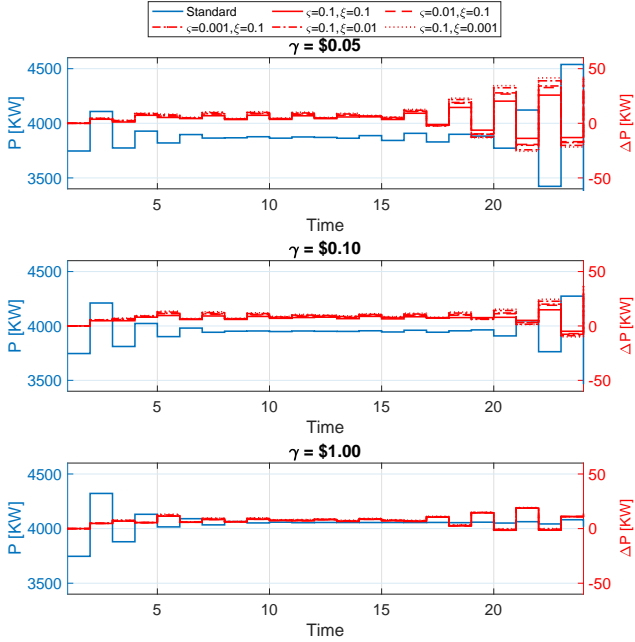}

\caption{\textcolor{black}{Optimal power dispatch under the standard MDP in~\eqref{base_mdp} (blue) and the difference (denoted as $\Delta P$) in the power consumption under the hybrid model ($\eta=0$) in~\eqref{hybrid_final} for different values of cost penalty $\gamma$.}}
\label{fig:hybrid2}
\end{figure}

This section studies the performance and solution quality attained with the analytical control policies derived in Theorems~\ref{theorem1}--\ref{theorem3}. We implement the hybrid model and use it to obtain the stochastic and distributionally robust solutions by setting $\eta = 1.00$ and $\eta = 0.00$, respectively. For the mean and variance bounds in \eqref{eq:ambiguity_set_mean} and \eqref{eq:ambiguity_set_var}, we set the values of parameters $\xi=0.001$ and $\varsigma=0.1$. Table~\ref{table_hybrid} summarizes the cost performance of all control policies for different values of $\eta$ and $\gamma$ and Fig.~\ref{fig:hybrid} itemizes the TCL ensemble power dispatch\footnote{\textcolor{black}{Here and in the following discussions, the power dispatch is recovered from the MDP solution as $p_{t}= \sum_{\beta \in \mathcal{A}} p^{\beta,rated} \rho_{t}^{\beta},\forall t \in \mathcal{T}$, where $p^{\beta,rated}$ is the rated power at each state and $\rho^{\beta}_{t}$ is the MDP solution.}} for selected values of $\eta$. As expected, the solution cost decreases as the value of parameter $\eta$ increases, i.e. distributional robustness and the ability to accommodate high-fidelity assumptions on the underlying uncertainty come at a modest increase in the operating cost. However, the cost increases also depend on the value of chosen cost penalty $\gamma$. As $\gamma$ increases, so does the cost difference between the stochastic and distributionally robust solutions. In terms of the power dispatch displayed in Fig.~\ref{fig:hybrid}, internalizing the uncertainty on transition probabilities tends to increase the flexibility of the TCL ensemble\footnote{\textcolor{black}{In this case study, the term flexibility refers to the difference between the default power consumption and the power consumption with one of the proposed MDP solutions.}} relative to the flexibility that can be extracted from the TCL ensemble relative to the standard MDP solution. In turn, the amount of this extra flexibility ($\Delta P$ in Fig.~\ref{fig:hybrid}) depends on the time period and on the value of cost penalty $\gamma$. The greater this cost penalty, the less flexibility can be extracted from the TCL ensemble. \textcolor{black}{We further evaluate the cost performance of the analytical control policies in the distributionally robust case ($\eta=0$) for different mean and variance bounds by varying parameters $\xi$ and $\varsigma$ in Table~\ref{table_hybrid_2} and
Fig.~\ref{fig:hybrid2}. %We set $\eta=0.00$ to fully explore the effect of these parameters on the distributionally robust solutions.
It is observed that the solution cost increases with a decrease in values of $\xi$ and $\varsigma$, and the magnitude of this increase is greater for greater values of cost penalty $\gamma$. This is expected because decreasing the values of $\xi$ and $\varsigma$ expands the confidence bounds around the mean and variance, which increases the robustness of the solution and immunizes it against a more extreme worst-case distribution.} 

Notably, the computational time for the analytical control policies in Theorems \ref{theorem1}--\ref{theorem3} is less than 0.013 seconds in all numerical experiments discussed above.

\begin{table}[!t]
\centering
\scriptsize
\caption{\textcolor{black}{Cost Performance of Analytical Control Policies in the Distributionally Robust Case ($\eta=0$).}}
\begin{tabular}{ c|c|>{\centering\arraybackslash}p{1.15cm}|>{\centering\arraybackslash}p{1.15cm}|>{\centering\arraybackslash}p{1.15cm}}
\hline
\multicolumn{5}{c}{\textcolor{black}{Objective function, \$}} \\
\hline
{\textcolor{black}{\multirow{2}{*}{$\gamma (\$)$}}} &\textcolor{black}{\textcolor{black}{\multirow{2}{*}{Parameter $\varsigma$}}} & \multicolumn{3}{c}{\textcolor{black}{Parameter $\xi$}} \\
{} & &\multicolumn{1}{c}{\textcolor{black}{0.1}} &\multicolumn{1}{c}{\textcolor{black}{0.01}} &\multicolumn{1}{c}{\textcolor{black}{0.001}} \\
\hline
{}  &\textcolor{black}{\multirow{2}{*}{0.1}} &\boldsymbol{\textcolor{black}{\multirow{2}{*}{2785.34$^*$}}} &\textcolor{black}{2786.25 (0.035\% $\uparrow$)} &\textcolor{black}{2787.04 (0.061\% $\uparrow$)} \\
\textcolor{black}{\multirow{2}{*}{0.05}} &\textcolor{black}{\multirow{2}{*}{0.10}} &\textcolor{black}{2785.88 (0.019\% $\uparrow$)} &\textcolor{black}{2786.88 (0.055\% $\uparrow$)} &\textcolor{black}{2787.75 (0.086\% $\uparrow$)} \\
{}  &\textcolor{black}{\multirow{2}{*}{0.001}} &\textcolor{black}{2786.35 (0.036\% $\uparrow$)} &\textcolor{black}{2787.42 (0.074\% $\uparrow$)} &\textcolor{black}{2788.36 (0.108\% $\uparrow$)} \\
\hline
{}  &\textcolor{black}{\multirow{2}{*}{0.1}} &\boldsymbol{\textcolor{black}{\multirow{2}{*}{2802.34$^*$}}} &\textcolor{black}{2803.81 (0.052\% $\uparrow$)} &\textcolor{black}{2805.10 (0.098\% $\uparrow$)} \\
\textcolor{black}{\multirow{2}{*}{0.10}} &\textcolor{black}{\multirow{2}{*}{0.01}} &\textcolor{black}{2803.16 (0.029\% $\uparrow$)} &\textcolor{black}{2804.77 (0.086\% $\uparrow$)} &\textcolor{black}{2806.17 (0.136\% $\uparrow$)} \\
{}  &\textcolor{black}{\multirow{2}{*}{0.001}} &\textcolor{black}{2803.87 (0.054\% $\uparrow$)} &\textcolor{black}{2805.59 (0.115\% $\uparrow$)} &\textcolor{black}{2807.09 (0.169\% $\uparrow$)} \\
\hline
{}  &\textcolor{black}{\multirow{2}{*}{0.1}} &\boldsymbol{\textcolor{black}{\multirow{2}{*}{2861.87$^*$}}}  &\textcolor{black}{2874.05 (0.425\% $\uparrow$)} &\textcolor{black}{2884.72 (0.798\% $\uparrow$)} \\
\textcolor{black}{\multirow{2}{*}{1.00}} &\textcolor{black}{\multirow{2}{*}{0.01}} &\textcolor{black}{2868.21 (0.221\% $\uparrow$)} &\textcolor{black}{2881.42 (0.683\% $\uparrow$)} &\textcolor{black}{2892.97 (1.086\% $\uparrow$)} \\
{}  &\textcolor{black}{\multirow{2}{*}{0.001}} &\textcolor{black}{2873.63 (0.410\% $\uparrow$)} &\textcolor{black}{2887.71 (0.902\% $\uparrow$)} &\textcolor{black}{2900.01 (1.332\% $\uparrow$)} \\
\hline 
\end{tabular}
\begin{flushleft}
\textcolor{black}{\boldsymbol{$\qquad \quad \ ^*$} Bold numbers are reference values.}
\end{flushleft}
\label{table_hybrid_2}
\end{table}

\subsection{Numerical Control Policies}

%\begin{table}[!b]
%\centering
%\scriptsize
%\caption{Cost Performance of Numerical Control Policies.}
%\vspace{3pt}
%\begin{tabular}{ c|c|c|c }
%\hline
%\multicolumn{4}{c}{Solution Cost (Objective function), \$} \\
%\hline
%{}& {$\gamma$=\$0.05} & {$\gamma$=\$0.10} & {$\gamma$=\$1.00}\\
%\hline
%Standard & 2781.05 & 2795.47 & 2810.09 \\
%Moment-based & 2895.53 & 3139.94 & 3411.03  \\
%Wasserstein-based & 2836.59 & 2845.51 & 2902.16 \\
%\hline 
%\end{tabular}
%\label{table_moment}
%end{table}

\begin{table}[!t]
\centering
\scriptsize
\caption{\textcolor{black}{Cost Performance of the Moment-Based MDP.}}
\begin{tabular}{ c|c|>{\centering\arraybackslash}p{1.15cm}|>{\centering\arraybackslash}p{1.15cm}|>{\centering\arraybackslash}p{1.15cm}}
\hline
\multicolumn{5}{c}{\textcolor{black}{Objective function, \$}} \\
\hline
{\textcolor{black}{\multirow{2}{*}{$\gamma (\$)$}}}& \textcolor{black}{\multirow{2}{*}{Parameter $c$}}& \multicolumn{3}{c}{\textcolor{black}{Parameter $b$}} \\
{}& &\multicolumn{1}{c}{\textcolor{black}{0.05}} &\multicolumn{1}{c}{\textcolor{black}{0.10}} &\multicolumn{1}{c}{\textcolor{black}{0.20}} \\
\hline
{}  &\textcolor{black}{\multirow{2}{*}{1.5}} &\boldsymbol{\textcolor{black}{\multirow{2}{*}{2766.81$^*$}}} &\textcolor{black}{2840.93 (2.67\% $\uparrow$)} &\textcolor{black}{2843.35 (2.76\% $\uparrow$)} \\
\textcolor{black}{\multirow{2}{*}{0.05}} &\textcolor{black}{\multirow{2}{*}{2.0}} &\textcolor{black}{2805.83 (1.41\% $\uparrow$)} &\textcolor{black}{2883.15 (4.20\% $\uparrow$)} &\textcolor{black}{2886.11 (4.31\% $\uparrow$)} \\
{}  &\textcolor{black}{\multirow{2}{*}{3.0}} &\textcolor{black}{2812.71 (1.65\% $\uparrow$)} &\textcolor{black}{2891.76 (4.51\% $\uparrow$)} &\textcolor{black}{2895.53 (4.65\% $\uparrow$)} \\
\hline
{}  &\textcolor{black}{\multirow{2}{*}{1.5}} &\boldsymbol{\textcolor{black}{\multirow{2}{*}{2976.63$^*$}}} &\textcolor{black}{3074.05 (3.27\% $\uparrow$)} &\textcolor{black}{3077.51 (3.38\% $\uparrow$)} \\
\textcolor{black}{\multirow{2}{*}{0.10}} &\textcolor{black}{\multirow{2}{*}{2.0}} &\textcolor{black}{3021.09 (1.49\% $\uparrow$)} &\textcolor{black}{3121.05 (4.85\% $\uparrow$)} &\textcolor{black}{3126.17 (5.02\% $\uparrow$)} \\
{}  &\textcolor{black}{\multirow{2}{*}{3.0}} &\textcolor{black}{3032.11 (1.86\% $\uparrow$)} &\textcolor{black}{3133.77 (5.27\% $\uparrow$)} &\textcolor{black}{3139.94 (5.48\% $\uparrow$)} \\
\hline
{}  &\textcolor{black}{\multirow{2}{*}{1.5}} &\boldsymbol{\textcolor{black}{\multirow{2}{*}{3179.43$^*$}}} &\textcolor{black}{3295.54 (3.65\% $\uparrow$)} &\textcolor{black}{3303.64 (3.90\% $\uparrow$)} \\
\textcolor{black}{\multirow{2}{*}{1.00}} &\textcolor{black}{\multirow{2}{*}{2.0}} &\textcolor{black}{3262.48 (2.61\% $\uparrow$)} &\textcolor{black}{3382.84 (6.39\% $\uparrow$)} &\textcolor{black}{3391.76 (6.67\% $\uparrow$)} \\
{}  &\textcolor{black}{\multirow{2}{*}{3.0}} &\textcolor{black}{3279.11 (3.13\% $\uparrow$)} &\textcolor{black}{3401.34 (6.97\% $\uparrow$)} &\textcolor{black}{3411.03 (7.28\% $\uparrow$)} \\
\hline
\end{tabular}
\begin{flushleft}
\textcolor{black}{\boldsymbol{$\qquad \quad \ ^*$} Bold numbers are reference values.}
\end{flushleft}
\label{table_moment_2}
\end{table}

\begin{table}[!t]
\centering
\scriptsize
\caption{\textcolor{black}{Cost Performance of the Wasserstein-Based MDP.}}

\begin{tabular}{ c|c|c|c }
\hline
\multicolumn{4}{c}{\textcolor{black}{Objective function, \$}} \\
\hline
{\textcolor{black}{$\gamma (\$)$}}& {\textcolor{black}{$\psi$=0.5}} & {\textcolor{black}{$\psi$=1.0}} & {\textcolor{black}{$\psi$=2.0}}\\
\hline
\textcolor{black}{0.05} & \boldsymbol{\textcolor{black}{2808.13$^*$}} & \textcolor{black}{2818.60 (0.37\% $\uparrow$)} & \textcolor{black}{2836.59 (1.01\% $\uparrow$)} \\
\textcolor{black}{0.10} & \boldsymbol{\textcolor{black}{2814.84$^*$}} & \textcolor{black}{2826.33 (0.40\% $\uparrow$)} & \textcolor{black}{2845.51 (1.08\% $\uparrow$)} \\
\textcolor{black}{1.00} & \boldsymbol{\textcolor{black}{2852.42$^*$}} & \textcolor{black}{2871.03 (0.65\% $\uparrow$)} & \textcolor{black}{2902.16 (1.74\% $\uparrow$)} \\
\hline 
\end{tabular}
\begin{flushleft}
\textcolor{black}{\boldsymbol{$\qquad \qquad^*$} Bold numbers are reference values.}
\end{flushleft}
\label{table_wass2}
\end{table}

\begin{figure}[!t]
\centering
\includegraphics[height=9cm,width=\columnwidth]{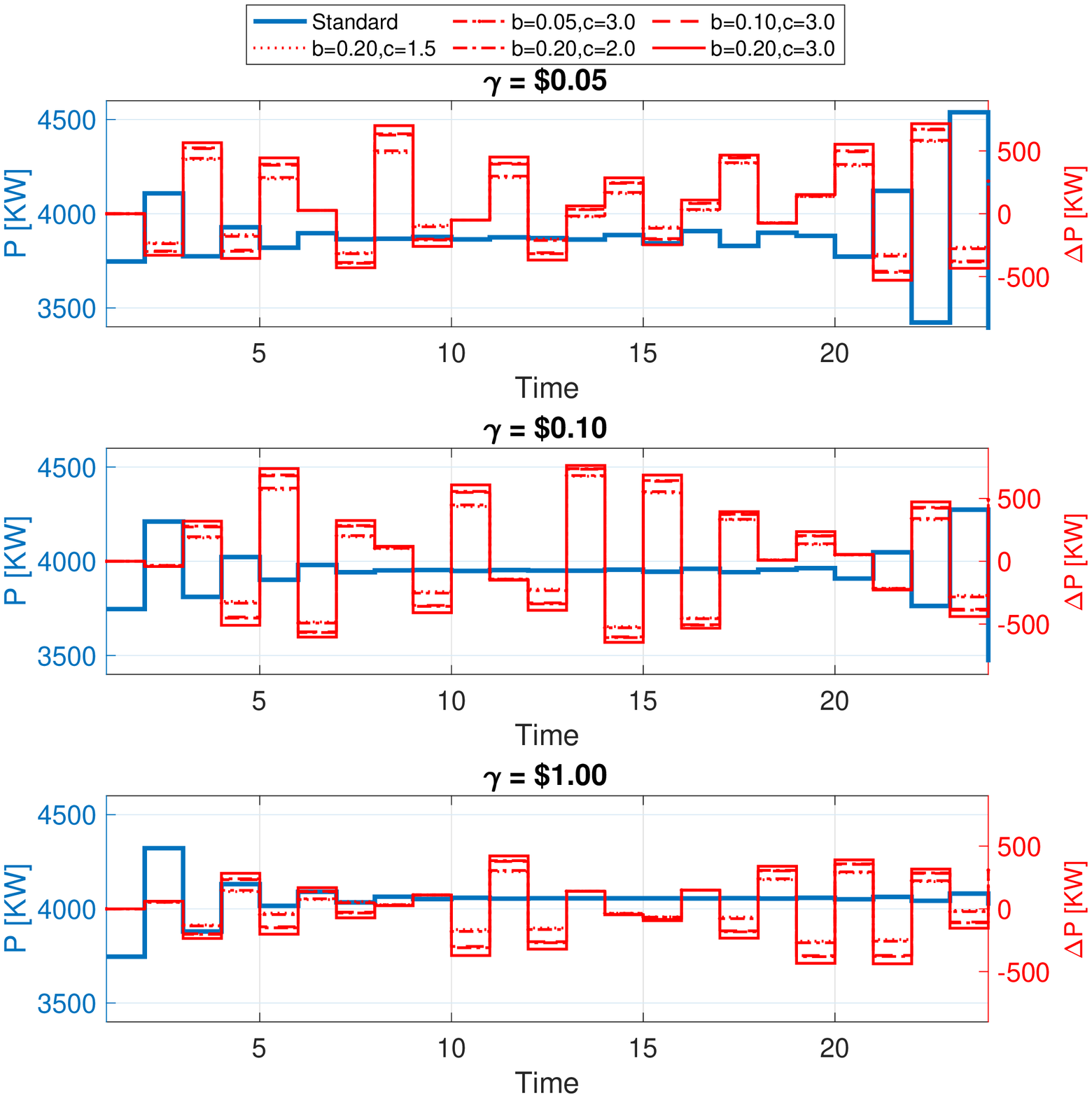}

\caption{Optimal power dispatch under the standard MDP in~\eqref{base_mdp} (blue) and the difference (denoted as $\Delta P$) in the power consumption under the moment-based distributionally robust MDP in~\eqref{moment_final} (red) for different values of cost penalty $\gamma$. }
\label{fig:moment}
\end{figure}

\begin{figure}[!t]
\centering
\includegraphics[height=9cm,width=\columnwidth]{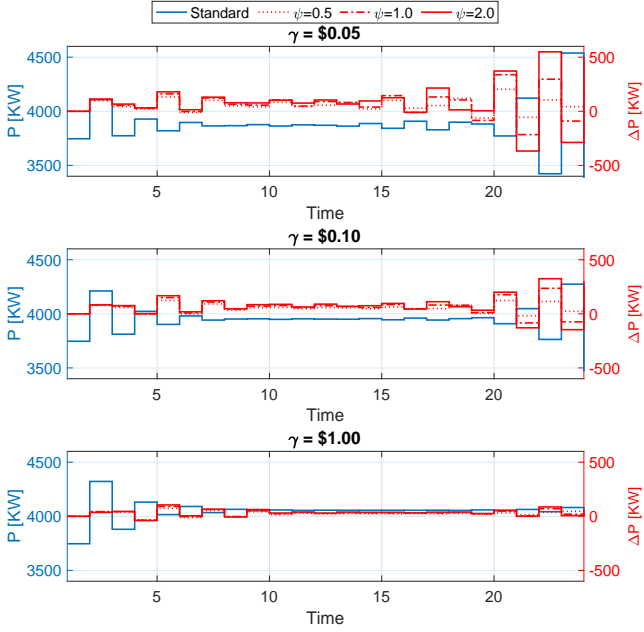}

\caption{Optimal power dispatch under the standard MDP in~\eqref{base_mdp} (blue) and the difference (denoted as $\Delta P$) in the power consumption under the Wasserstein-based distributionally robust MDP in~\eqref{wass_3} (red) for different values of cost penalty $\gamma$.
}
\label{fig:wass}
\end{figure}

This section compares the cost and dispatch performance of distributionally robust solutions obtained using numerical control polices described in Section~\ref{Numerical Control}. 
%We set $b=0.2$ and $c=3$ in \eqref{moment_constraints} for the moment-based ambiguous uncertainty set and $\psi=2$ in \eqref{eq:wass_ball} for the Wasserstein-based ambiguous uncertainty set. Table~\ref{table_moment} presents the solution cost for different values of parameter $\gamma$ and Figures~\ref{fig:moment} and \ref{fig:wass} compare the power dispatch of the TCL ensembles under different ambiguity sets relative to the standard MDP formulation.
\textcolor{black}{
%We define multiple values for $b$ and $c$ in \eqref{moment_constraints} for the moment-based ambiguous uncertainty set and $\psi$ in \eqref{eq:wass_ball} for the Wasserstein-based ambiguous uncertainty set.
Tables~\ref{table_moment_2} and \ref{table_wass2} present the solution cost for different values of parameter $\gamma$ and Figures~\ref{fig:moment} and \ref{fig:wass} compare the power dispatch of the TCL ensembles under moment- and Wasserstein-based ambiguity sets relative to the standard MDP formulation for different values of parameters $b$ and $c$ in \eqref{moment_constraints} and $\psi$ in \eqref{eq:wass_ball}.}
Naturally, the solution cost increases for greater values of cost penalty $\gamma$. Under both the moment- and Wasserstein-based ambiguity  sets, the solution cost increases relative to the standard MDP and analytical control policies in Table~\ref{table_hybrid}. These operating cost increases are expected, because using the ambiguous uncertainty sets makes it possible to better accommodate empirical observations, i.e. without assuming normally distributed errors on transition probabilities. %Notably, the Wasserstein-based ambiguous uncertainty set leads to lower operating costs than the moment-based approach, regardless of the chosen value for cost penalty $\gamma$. These cost savings are observed since the uncertainty set defined in \eqref{eq:wass_ball} is used to control solution conservatism by adjusting radius $\psi$.
In terms of the power dispatch, the moment-based approach leads to more volatile dispatch decisions for all values of cost penalty $\gamma$ than in the Wasserstein-based case. Relative to the standard case, both the moment- and Wasserstein-based cases tend to increase the overall power flexibility ($\Delta P$ in Fig.~\ref{fig:moment} and \ref{fig:wass}) extracted from the TCL ensemble over 24 hours. 
\textcolor{black}{Similar to the analytical control policies in Section~\ref{Case Study}.\ref{Case Study:Analytical}, we analyze the effects of confidence parameters on the cost performance of moment-based and Wasserstein-based methods. For the moment-based method, as presented in Table~\ref{table_moment_2}, the solution cost increases as the confidence region around the first and second-order moments widens by changing the values of parameters $b$ and $c$. Fig.~\ref{fig:moment} displays the effect of varying $b$ and $c$ on the power dispatch of the TCL ensembles, where more inter-temporal fluctuations are observed for greater values of parameters $b$ and $c$. In addition, in the Wasserstein-based method, we observe an increase in solution costs, see Table~\ref{table_wass2}, and power dispatch fluctuations, see Fig.~\ref{fig:wass}, as radius $\psi$  around the nominal distribution increases. }

The average computational times for the moment- and Wasserstein-based cases are 18.2 and 44.5 seconds, which is significantly greater than for the analytical control policies. 

%\begin{figure}[!t]
%\centering
%\includegraphics[height=9cm,width=\columnwidth]{Power_moment_new.eps}
%\vspace{-7mm}
%\caption{}
%\label{fig:moment2}
%\end{figure}

%\begin{figure}[!t]
%\centering
%\includegraphics[height=9cm,width=\columnwidth]{Power_wass2.eps}
%\vspace{-7mm}
%\caption{Optimal power dispatch under the standard MDP in~\eqref{base_mdp} (blue) and the difference (denoted as $\Delta P$) in the power consumption under the Wasserstein-based distributionally robust MDP in~\eqref{wass_3} (red) for different values of cost penalty $\gamma$.
%}
%\label{fig:wass2}
%\end{figure}

\section{Conclusion}
This paper describes analytical and numerical approaches to internalize the uncertainty dynamics of TCL ensembles in the Markov Decision Problem using stochastic and distributionally robust optimization. The stochastic and distributionally robust control policies are derived under mild assumptions on the underlying uncertainty and can be implemented in a computationally efficient manner. On the other hand, allowing for computationally demanding numerical control policies allows for better fitting empirical data, thus producing more accurate control policies and reducing data requirements for MDP problems. Our case study demonstrates that both the analytical and numerical control policies improve the accuracy of computing dispatch flexibility that can be extracted from the TCL ensemble relative to the standard MDP optimization, while minimizing the level of discomfort incurred to TCL users. 
\textcolor{black}{Among different methods to accommodate the uncertainty in empirical measurements of TCL ensemble, we find that robust methods have more exogenous parameters that can be leveraged to intelligently trade-off solution cost and robustness. Although these exogenous parameters vary for the moment- and Wasserstein-based approaches, our numerical results demonstrate that they can be tuned in each case to achieve a comparable cost performance, thus allowing for distributionally robust decision-making in applications with different data availability.}

\appendices
\section{Proofs of Theorems \ref{theorem0}--\ref{theorem3}} \label{sec:exp_zero_mean}
% This section presents the proof of Theorems 1--4.
We follow a similar procedure to prove all Theorems 1--4, where theorem-specific terms are denoted as $\mathcal{Z}$. The value of $\mathcal{Z}$ for each theorem is derived at the end of this appendix. For each theorem, given its respective MDP optimization, we can write the following Bellman equation $\forall t$ and $\forall \beta$: 
\allowdisplaybreaks
\begin{align}
\begin{split}
& \!\!\!\frac{1}{\gamma}\varphi_{t}^{\beta} = \frac{1}{\gamma} \underset{\substack{\mathcal{P}}}{\text{min}} \Big(\!\!-U_{t}^{\beta}\! + \mathbb{E}_{\mathcal{P}^{\alpha\beta}} \! \Big[\gamma \log \frac{\mathcal{P}_{t}^{\alpha\beta}} {{\overline{\mathcal{P}}^{\alpha\beta}}} + \mathcal{Z} + \varphi_{t+1}^{\alpha}\Big] \!\Big), \label{bellmen_1} 
\end{split} 
\end{align}
where $\varphi_{t}^{\beta}$ is the value function at the present state $\beta$, $\varphi_{t+1}^{\alpha}$ is the value function from the next state $\alpha$ and $\mathcal{Z}$ represents a theorem-specific term for any possible transition probability uncertainty. Introducing the auxiliary (desirability) function $z^{\beta} = \text{exp}(-\varphi_{t}^{\beta}/\gamma)$ in \eqref{bellmen_1} leads to:

\begin{align}
\begin{split}
& -\!\!\text{log}(\!z_{t}^{\beta})\! =\! \frac{1}{\gamma} \underset{\substack{\mathcal{P}}}{\text{min}} \Big(\!\!-\!U_{t}^{\beta}\!\! +\! \gamma \mathbb{E}_{\mathcal{P}^{\alpha\beta}}\! \Big[\! \log\! \frac{\mathcal{P}_{t}^{\alpha\beta}} {{\overline{\mathcal{P}}^{\alpha\beta}}} \! + \!\mathcal{Z} \!-\! \text{log}(\!z_{t+1}^{\alpha})\! \Big] \Big) = 
\end{split}\nonumber\\
\begin{split}
&\frac{1}{\gamma} \underset{\substack{\mathcal{P}}}{\text{min}} \Big(-U_{t}^{\beta} + \gamma \mathbb{E}_{\mathcal{P}^{\alpha\beta}} \Big[ \log \frac{\mathcal{P}_{t}^{\alpha\beta}} {{\overline{\mathcal{P}}^{\alpha\beta}}z_{t+1}^{\alpha}\text{exp}(-\mathcal{Z} )} \Big] \Big). \label{bellmen_5}
\end{split}
\end{align}
Next, the right-hand side of~\eqref{bellmen_5} is normalized using $\mathcal{G}^{\beta}(z)=\sum_{\alpha}\overline{\mathcal{P}}^{\alpha\beta}z_{t+1}^{\alpha}\text{exp}(-\mathcal{Z})$, which results in:

\begin{align}
\begin{split}
& -\text{log}(z_{t}^{\beta}) = \frac{1}{\gamma} \underset{\substack{\mathcal{P}}}{\text{min}} \bigg(-U_{t}^{\beta} + \gamma \mathbb{E}_{\mathcal{P}^{\alpha\beta}} \bigg[ \log\\& \frac{\mathcal{P}_{t}^{\alpha\beta}\mathcal{G}^{\beta}(z)} {\sum_{\alpha}\overline{\mathcal{P}}^{\alpha \beta}z_{t+1}^{\alpha}\text{exp}(-\mathcal{Z}) \mathcal{G}^{\beta}(z)} \bigg] \bigg) = \bigg(\frac{-U_{t}^{\beta}}{\gamma} + \underset{\substack{\mathcal{P}}}{\text{min}} 
\end{split} \nonumber \\
\begin{split}
& \KL\! \bigg[\mathcal{P}_{t}^{\alpha\beta} \bigg\Vert \frac{\sum_{\alpha}\overline{\mathcal{P}}^{\alpha \beta}z_{t+1}^{\alpha}\text{exp}(-\mathcal{Z})}{\mathcal{G}^{\beta}(z)} \bigg] - \text{log}\mathcal{G}^{\beta}(z)\! \bigg), \label{bellmen_9} 
\end{split}
\end{align}
where $\KL\big[\cdot|| \cdot \big]$ denotes the KL-divergence. The optimal policy is achieved when $\KL$ term in~\eqref{bellmen_9} is minimal, i.e. equal to zero. Since the zero value of the KL divergence is achieved when both distributions are identical, we obtain the condition for the optimal policy as:
\begin{align}
&\mathcal{P}_{t}^{\alpha \beta} = \frac{\overline{\mathcal{P}}^{\alpha \beta}z_{t+1}^{\alpha}\text{exp} (-\mathcal{Z})}{\mathcal{G}^{\beta}(z)} % = \frac{\overline{\mathcal{P}}^{\alpha \beta}z^{\alpha}\text{exp}(-\mathcal{Z})}{\sum_{\alpha}\overline{\mathcal{P}}^{\alpha \beta}z^{\alpha}\text{exp}(-\mathcal{Z})} 
\label{optimal_policy_1}
\end{align}
Using the optimal policy in \eqref{optimal_policy_1} and recalling that $\mathcal{G}^{\beta}(z)=\sum_{\alpha}\overline{\mathcal{P}}^{\alpha\beta}z_{t+1}^{\alpha}\text{exp}(-\mathcal{Z})$,  the Bellman equation in \eqref{bellmen_9} can be recast as:
\begin{align}
\begin{split}
& -\text{log}(z_{t}^{\beta}) = \{{-U_{t}^{\beta}}/{\gamma} -\text{log}\mathcal{G}^{\beta}(z) \} \label{bellmen_reduced_1} 
\end{split} \\  
\begin{split}
& \text{log}(z_{t}^{\beta}) = \Big\{{U_{t}^{\beta}}/{\gamma} + \text{log} \Big[\sum_{\alpha}\overline{\mathcal{P}}^{\alpha \beta}z_{t+1}^{\alpha} \text{exp}(-\mathcal{Z})\Big] \Big\} \label{bellmen_reduced_2} 
\end{split}
\end{align}
Exponentiating~\eqref{bellmen_reduced_2} leads to:
\begin{align}
&z_{t}^{\beta} = \text{exp}\Big({U_{t}^{\beta}}/{\gamma}\Big) \sum_{\alpha}\overline{\mathcal{P}}^{\alpha \beta}z_{t+1}^{\alpha} \text{exp} (-\mathcal{Z}) \label{bellmen_reduced_3}.
\end{align}
Since the value of $\mathcal{Z}$ varies for Theorems 1-4, we derive theorem-specific results for each case below.

\subsection{Standard Formulation in Theorem 1} 
The standard model ignores the uncertainty of transition probabilities, which leads to:
\begin{align}
&\mathcal{Z}^{S} := \mathcal{Z} = 0. \label{standard_Z}
\end{align}
Accordingly, using \eqref{standard_Z} returns the following optimal policy:
\begin{align}
&\mathcal{P}_{t}^{\alpha \beta} = \frac{\overline{\mathcal{P}}^{\alpha \beta}z_{t+1}^{\alpha}}{\sum_{\alpha}\overline{\mathcal{P}}^{\alpha \beta}z_{t+1}^{\alpha}} .
\end{align}

\subsection{Stochastic Formulation in Theorem 2} \label{Appendix:stochastic}
The value of $\mathcal{Z}$ for the stochastic model follows from \eqref{uncer_obj_final} as:
\begin{align}
&\mathcal{Z}^{E} := \mathcal{Z} = \frac{(\gamma \sigma^2)}{2({\overline{\mathcal{P}}^{\alpha\beta}})^2} \label{stochastic_Z} 
\end{align}
Accordingly, using \eqref{stochastic_Z} returns the following optimal policy:
\begin{align} \label{eq:z_stoch}
&\mathcal{P}_{t}^{\alpha \beta} = \frac{\overline{\mathcal{P}}^{\alpha \beta}z_{t+1}^{\alpha}\text{exp}\Big(\frac{-\gamma \sigma^2} {2({\overline{\mathcal{P}}^{\alpha\beta}})^2}\Big)}{\sum_{\alpha}\overline{\mathcal{P}}^{\alpha \beta}z_{t+1}^{\alpha}\text{exp}\Big(\frac{-\gamma \sigma^2} {2({\overline{\mathcal{P}}^{\alpha\beta}})^2}\Big)}.
\end{align}

\subsection{Distributionally Robust Formulation in Theorem 3}
The value of $\mathcal{Z}$ for the distributionally robust formulation follows from \eqref{chi_obj} as:
\begin{align}
& \mathcal{Z}^{WC} :=\mathcal{Z} = \frac{(\gamma \overline{\hat{\zeta}})}{2(\underline{\Gamma})^2} \label{DRO_Z}  
\end{align}
Accordingly, using \eqref{DRO_Z} returns the following optimal policy:
\begin{align} \label{eq:z_dro}
&\mathcal{P}_{t}^{\alpha \beta} = \frac{\underline{\Gamma}z_{t+1}^{\alpha}\text{exp}\Big(\frac{-\gamma \overline{\hat{\zeta}}} {2(\underline{\Gamma})^2}\Big)}{\sum_{\alpha}\underline{\Gamma}z_{t+1}^{\alpha}\text{exp}\Big(\frac{-\gamma \overline{\hat{\zeta}}} {2(\underline{\Gamma})^2}\Big)}.
\end{align}
where $\overline{\mathcal{P}}^{\alpha\beta}$ is replaced by its bound $\underline{\Gamma}$ from the set $\mathbb{D}$ to obtain the worst-case distribution.

\subsection{Hybrid Model in Theorem 4}
Using \eqref{eq:z_stoch} and \eqref{eq:z_dro}, the hybrid optimal policy follows as:
\begin{align}
\begin{split}
&\hspace{-5mm}\mathcal{P}_{t}^{\alpha \beta} = (1-\eta)\frac{\underline{\Gamma}z_{t+1}^{\alpha}\text{exp}\Big(\frac{-\gamma \overline{\hat{\zeta}}} {2(\underline{\Gamma})^2}\Big)}{\sum_{\alpha}\underline{\Gamma}z_{t+1}^{\alpha}\text{exp}\Big(\frac{-\gamma \overline{\hat{\zeta}}} {2(\underline{\Gamma})^2}\Big)} \\& \qquad \qquad \qquad + \eta \frac{\overline{\mathcal{P}}^{\alpha \beta}z_{t+1}^{\alpha}\text{exp}\Big(\frac{-\gamma \sigma^2} {2({\overline{\mathcal{P}}^{\alpha\beta}})^2}\Big)}{\sum_{\alpha}\overline{\mathcal{P}}^{\alpha \beta}z_{t+1}^{\alpha}\text{exp}\Big(\frac{-\gamma \sigma^2} {2({\overline{\mathcal{P}}^{\alpha\beta}})^2}\Big)}  
\end{split}
\end{align}
where $0\leq\eta\leq1$.

\bibliographystyle{IEEEtran}
\bibliography{ref.bib}

\end{document}